\documentclass[twocolumn]{aastex631}

\shorttitle{Rocky histories}
\shortauthors{Scora et al.}

\graphicspath{{./}{figures/}}

\begin{document}

\title{Rocky histories: The effect of high excitations on the formation of rocky planets}

\correspondingauthor{Jennifer Scora}
\email{scora@astro.utoronto.ca}

\author{Jennifer Scora}
\affil{David A. Dunlap Department of Astronomy and Astrophysics \\
University of Toronto \\
Toronto, ON, Canada, M5S 3H4}

\author[0000-0003-3993-4030]{Diana Valencia}
\affiliation{ Centre for Planetary Sciences \\
University of Toronto \\
1265 Military Trail \\
Toronto, ON, M1C 1A4, Canada}

\author{Alessandro Morbidelli}
\affil{Laboratoire Lagrange, UMR7293, Universite de Nice Sophia-Antipolis, CNRS, \\
Observatoire de la Cote d'Azur. Boulevard de l'Observatoire, 06304 Nice Cedex 4, France}

\author[0000-0002-4952-9007]{Seth Jacobson}
\affil{Department of Earth and Environmental Sciences \\
Michigan State University \\
East Lansing, MI 48824, USA}

\begin{abstract}

Rocky planets both in and outside of our solar system are observed to have a range of core-mass fractions (CMFs). Imperfect collisions can preferentially strip mantle material from a planet, changing its CMF, and are therefore thought to be the most likely cause of this observed CMF variation. However, previous work that implements these collisions into N-body simulations of planet formation has struggled to reliably form high CMF super-Earths. In this work, we specify our initial conditions and simulation parameters to maximize the prevalence of high-energy, CMF-changing collisions in order to form planets with highly diverse CMFs. High-energy collisions have a large $v_{imp}/v_{esc}$ ratio, so we maximize this ratio by starting simulations with high-eccentricity and inclination disks to increase the difference in their orbital velocities, maximizing $v_{imp}$. Additionally, we minimize $v_{esc}$ by starting with small embryos. The final planets undergo more high-energy, debris-producing collisions, and experience significant CMF change over their formation. However, we find that a number of processes work together to average out the CMF of a planet over time, therefore we do not consistently form high-CMF, high mass planets. We do form high-CMF planets below 0.5 $M_{\oplus}$. Additionally, we find in these highly eccentric environments, loss of debris mass due to collisional grinding has a significant effect on final planet masses and CMFs, resulting in smaller planets and a higher average planet CMF. This work highlights the importance of improving measurements of high-density planets to better constrain their CMFs. 

\end{abstract}

\section{Introduction} \label{sec:intro}

There are more than a thousand observed exoplanets with radii less than $2 R_{\oplus}$. These low-mass planets are thought to be some of the most common planets in the galaxy \citep{Howard,2013Fressin}, thus as instrumentation improves this number will continue to increase. Planets below $1 R_{\oplus}$ (terrestrial planets) are rocky, but within the $1- 2 R_{\oplus}$ range, there exists three main families of planets: rocky and two types of volatile-rich planets. The former are thought to have silicate mantles and iron cores, similar to Earth, possibly with very tenuous atmospheres. The latter have a substantial amount of volatiles. These can be either sub-Neptunes with a gaseous envelope that considerably affects its radius \citep{2010Rogers}, or ocean/icy planets \citep{2007Valencia,2004Leger} 

This classification based on composition is useful because it divides the planets into categories that presumably had different formation pathways. In this work we focus on the formation of low-mass rocky planets: terrestrial planets and super-Earths. Unfortunately, it is not possible to distinguish rocky planets from volatile planets with single mass or radius measurements, because super-Earths and sub-Neptunes overlap in mass-radius space. Albeit, volatile planets are not predominant below $1.62 R_{\oplus}$ \citep{2015Rogers}, and $5 M_{\oplus}$ \citep{2020Otegi}.   

To select the rocky planets only, we consider exoplanets with precise mass and radius measurements that are below the rocky threshold radius (RTR). The RTR is the largest radius a rocky planet can have for its mass, which occurs when it has no iron. This threshold divides planets between those that require substantial amounts of volatiles (above the RTR), and those that are possibly but not necessarily rocky (below the RTR). This is because of degeneracy in composition, where a planet with the same mass and radius can match both a rocky model and a volatile model with a higher core-mass fraction and H/He envelope \citep{2010Rogers}, or substantial water  \citep{2007Valencia}.  Therefore, by assuming that all planets below the RTR are rocky we are considering the lowest core-mass fraction (CMF) for these planets. Any model with volatiles (i.e. sub-Neptune) would require a higher CMF to match the same radius and mass. 

\begin{figure}
    \centering
    \includegraphics[width=0.45\textwidth]{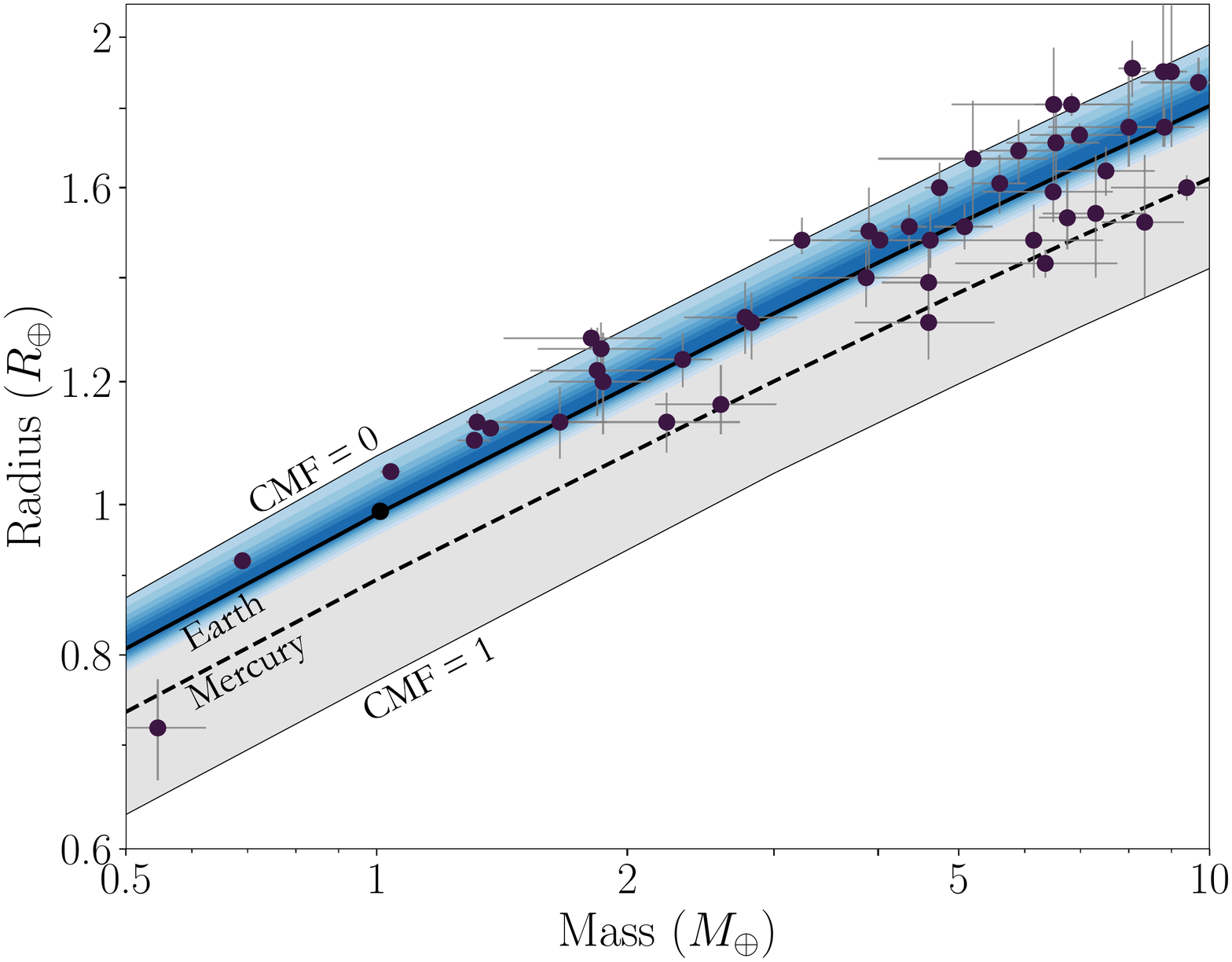}
    \caption{Observed rocky planets below 10 $M_{\oplus}$ with good mass and radius measurements taken from the NASA exoplanet archive \citep{NEA}, plotted over lines of constant CMF. The top line is a planet with no core (CMF = 0), the middle two lines are Earth (CMF = 0.326 \citep{2005Stacey}) and Mercury's (CMF $\sim$ 0.69 \citep{2013Hauck}) CMF respectively, and the bottom line is a planet made of only core (CMF = 1). The blue shaded area is the region that planets should occupy if they follow stellar compositional ratios, taken from \citep{2020Scora}. The planets span the range of possible CMFs, and have a maximum CMF close to Mercury's.}
    \label{fig:obs-super}
\end{figure}

Moving forward with this assumption, Figure \ref{fig:obs-super} shows that the sample of observed rocky planets span a wide range of CMFs, from planets with no cores at all to planets with very high CMFs. Super-Earths such as Kepler 105c \citep{2016Jontof-Hutter,2017Hadden}, Kepler 406b \citep{2016Morton}, K2-38b \citep{2020Toledo}, and Kepler 107c \citep{2021Schulze}, and recently discovered sub-Earth GJ~367 b \citep{Lam2021} all appear to have CMFs around 0.6 or higher. This variation in CMF does not just happen between systems, but from planet to planet within systems. Even the terrestrial planets in our solar system span a range of CMFs, from the Earth's CMF of 0.326 \citep{2005Stacey} to Mercury with the much higher CMF of 0.69 - 0.77 \citep{2013Hauck}. 

The formation of rocky planets with such a diversity of CMFs is still not fully understood. Rocky planets are often assumed to have similar compositions to their stars \citep{2015Dorn,2021Bonsor}. However, variations in stellar metallicities are not large enough to explain these large differences in CMF \citep{Plotnykov2020}. While it seems that at least some of these high CMF planets form around stars with higher iron abundances than the Sun, they are not enriched enough to form these high CMF planets \citep{2021Adibekyan}. Additionally, stellar abundances cannot explain the variation in CMF within a system, such as between the Earth and Mercury. Some new theories suggest a way that planetesimals closer to the Sun could be enriched in iron, contributing to Mercury's large core, but more work is needed to fit Mercury's other parameters (i.e. mass, eccentricity) \citep{2020Aguichine,2022Johansen}. The most plausible explanation that has frequently been put forth for Mercury's formation is that a giant collision (or many) stripped much of its silicate mantle away, increasing the CMF of the planet \citep{2007Benz}. Since a giant collisions phase is a part of both migration and in-situ super-Earth formation theories \citep{2014Raymond}, it is reasonable to consider this process of formation for both terrestrial planets and super-Earths. \citet{Marcus2009} and \citet{2022Reinhardt}, for example, have shown that individual collisions can generate a significant CMF increase.

\citet{2020Scora} explored how much CMF variation can be achieved in super-Earths due to giant collisions, and found that the maximum CMF planets were just short of Mercury's CMF. Those planets that did have high CMFs were rare, part of a low-likelihood tail in the distribution of planet CMFs. Most planets were within ~0.1 of the peak CMF (0.35), which was only slightly higher than the original CMF given to all embryos (0.33). The reason for the lack of spread in CMFs was that the disk was too dynamically cold; most collisions occurred at low ratios of impact to mutual escape velocities that only had minor effects on the planet's CMF. The types of collisions that increase a planet's CMF most, i.e. the more energetic ones that are able to strip more mantle material, occur rarely. When they do occur, often the debris created are simply reaccreted onto the same planet that they came from, resulting in an overall minimal change in composition. This has also been the case in a number of other studies that simulate rocky planet formation with imperfect accretion \citep{Chambers,Esteves2021}. Simulations of super-Earth formation struggle to consistently and commonly form high mass, high CMF bodies. \citet{2021Cambioni} found that they could only achieve high CMFs for planets with masses below $0.1 M_{\oplus}$.

In this study, we build on the work of \citet{2020Scora} by maximizing the number of energetic, composition-changing collisions that rocky planets experience as they form. Energetic collisions have high impact velocity to escape velocity ratios. To maximize this ratio, we decrease the mass of the embryos at the start of the simulation, decreasing their escape velocities. This is equivalent to starting our simulations earlier in the formation process, allowing for more time for those energetic collisions to accumulate throughout planet formation. Additionally, we ensure that the collisional impact velocities are higher by increasing the initial eccentricities and inclinations of the embryos, giving them high relative velocities. It is traditionally assumed that the initial eccentricities and inclinations are small, as in \citet{2020Scora}, because embryos form from a disk of gas and planetesimals. However, there is evidence that some super-Earths systems may have formed from planetary embryos on eccentric orbits. In fact, it is thought that as many as $\sim 30\%$ or more of super-Earth systems could have outer giant planet companions \citep{2018Zhu,2021Rosenthal}. Studies of the formation of the solar system have demonstrated that the movements of outer giant planets can excite the orbits of inner rocky bodies \citep{Walsh2011,2016Kaib}. \citet{2020Poon} simulated the evolution of super-Earths systems when perturbed by the motions of outer giant planets, and found that this could significantly increase the eccentricities of the inner planets to the point of causing collisions between previously stable super-Earths systems. Thus, high-eccentricity disks are plausible if we consider the influences of possible giant planets that have shaped the disk prior to the start of our simulations. 

This paper presents our simulations of rocky planets forming from small embryos in excited disks. Section \ref{sec:meth} details the methods and initial conditions used. Section \ref{sec:system} lays out the results of these simulations, and Sections \ref{sec:CMF-collhist} and \ref{sec:CMF-mass-coll} further explain our results. In Section \ref{sec:implications} and \ref{sec:disc} we discuss implications and our interpretations of the results and their impacts, and Section \ref{sec:conc} summarizes our conclusions.

\section{Modeling Formation with an N-body Gravitational Code} \label{sec:meth}

\begin{figure}
    \centering
    \includegraphics[width=0.48\textwidth]{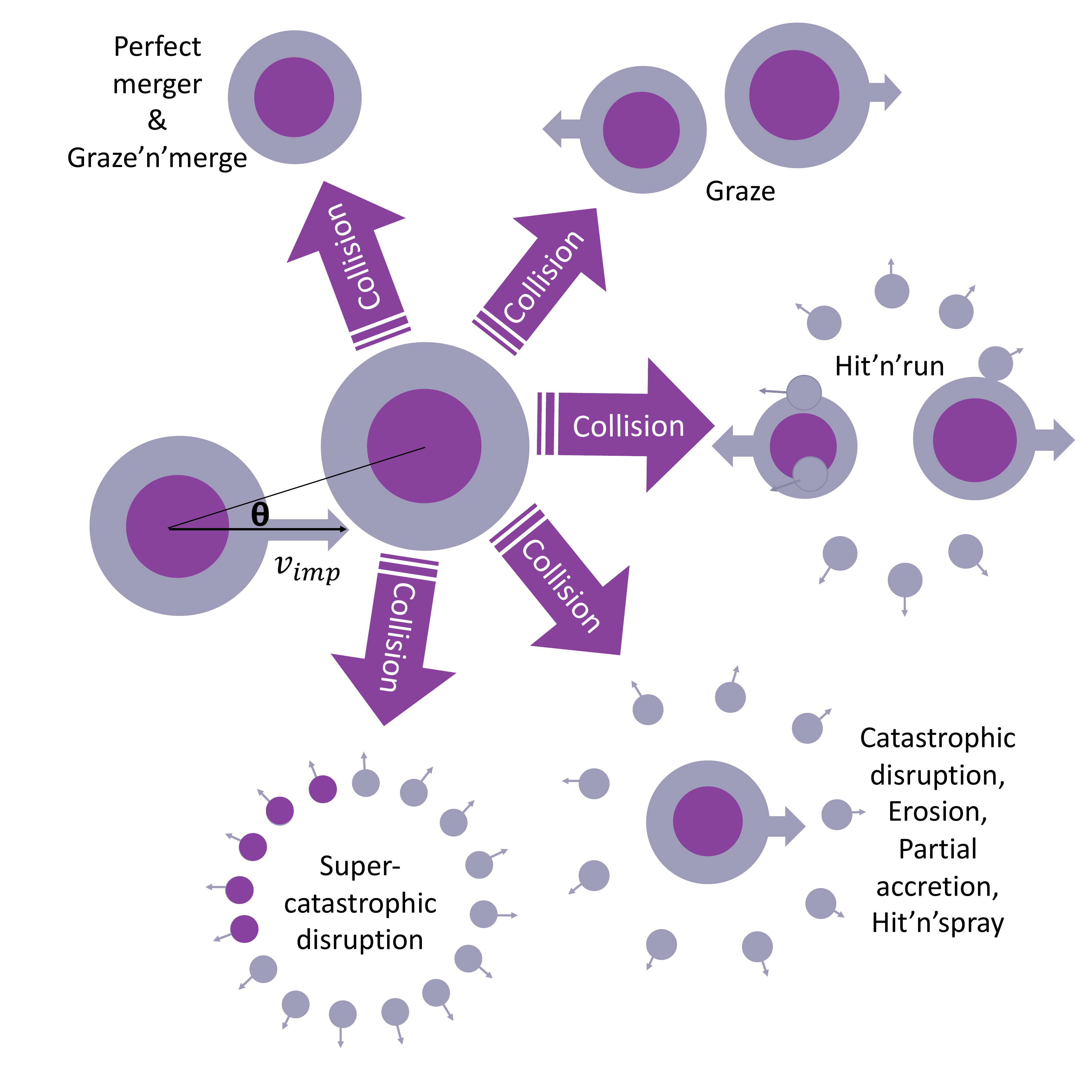}
    \caption{Visualization of the morphology and change in core-mass fraction for all collision types. Grey represents the mantle of a body, and purple its core. When two bodies collide with angle $\theta$ and impact velocity $v_{imp}$, there are nine different collision types assigned by the code (following \citet{Leinhardt2012} and \citet{Stewart}), but four main outcomes. Merger (top); a grazing collision with no real effect on the mass or CMF of the bodies (top right); a grazing impact that significantly strips one body, changing CMF and producing debris (right); an impact that leaves only one body intact and some debris (bottom right); a destructive impact that leaves only debris (bottom left).}
    \label{fig:coll-graphic}
\end{figure}

We simulate the formation of terrestrial super-Earths using a modified version of the gravitational N-body code \texttt{SyMBA} \citep{Duncan1998,2020Scora}. 
The code uses a multi-timestep symplectic integrator to efficiently resolve close encounters between bodies. It tracks the movement of particles within the disk based on gravitational interactions alone. Only embryos can gravitationally interact with each other; debris particles below a certain mass threshold ($0.04 M_{\oplus}$ in our simulations) have no gravitational effect on each other but do interact with the embryos. 

This code is an updated version that uses the analytic collision prescriptions of \citet{Leinhardt2012}, \citet{Stewart}, and \citet{Genda2012} to realistically calculate the outcomes of collisions between large bodies. Collisions are divided into nine different types based on their impact energy, mass ratio, and impact parameter, as depicted in Figure \ref{fig:coll-graphic}. There are four main outcomes. Super-catastrophic disruption collisions, the most destructive, leave only debris. Other disruptive collisions, such as erosive collisions, and one type of grazing collision, the hit'n'spray, leave one embryo remaining and some debris, either stripped from the target or the projectile. The other grazing collisions leave two embryos, and in the case of the hit'n'run some debris as well. Finally, both graze'n'merge and merge collisions are perfect mergers, leaving only one embryo. When debris are  generated in a collision, they are limited to a maximum of 38 particles and a minimum mass of $1.5 \times 10^{-5} M_{\oplus}$ each. See \citet{2020Scora} for a more detailed description of the collision treatment used in these simulations. 

The simulations also take into account the collisional grinding of debris into dust, and the subsequent loss of that dust due to radiation pressure pushing it away from the star \citep{2012Jackson} in a parameterized way. This process should happen in a few orbits \citep{2012Jackson}, which is a short enough timescale compared to the overall evolution that we approximate it to occur instantaneously. After each collision, the code removes a fraction of the debris mass created to simulate this effect. 

\subsection{Initial Conditions}\label{sec:IC}

\begin{table}
    \caption{Parameters for initial conditions of our simulation suite. The $e_{avg}$ and inc$_{avg}$ refer to the average eccentricity and inclination used in the Rayleigh distributions to assign values to the initial embryos.}
    \centering
    \begin{tabular}{ c | c | c | c }
         e$_{avg}$ & inc$_{avg}$ & debris loss & $\alpha$ \\
         \hline
         0.1 & 0.2 & 0$\%$ & 1.5 \\
         0.3 & 0.5 & 50$\%$ & 2.5 \\
         0.5 & & & \\
    \end{tabular}
    \label{tab:init-params}
\end{table}

\begin{figure*}
    \centering
    \plottwo{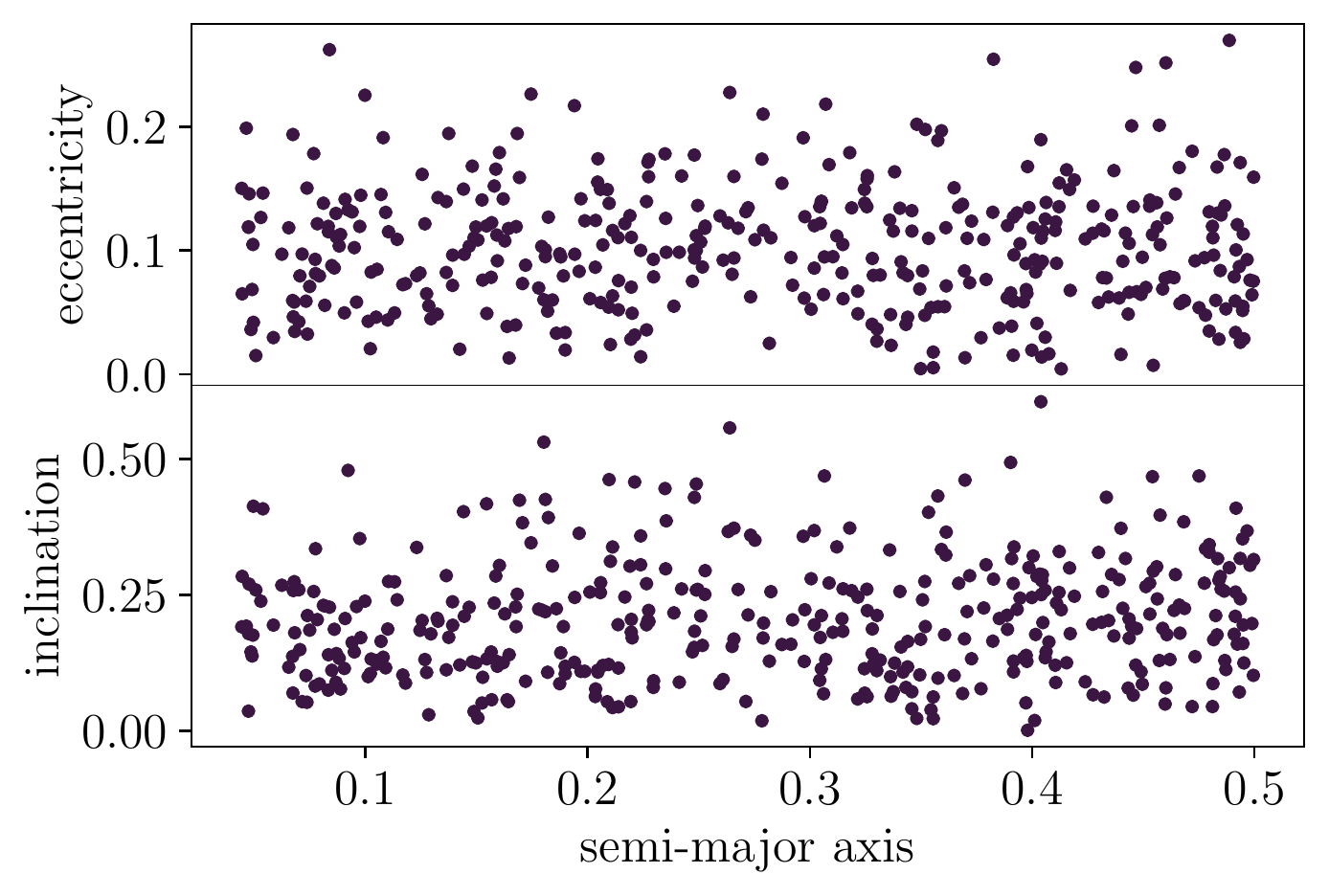}{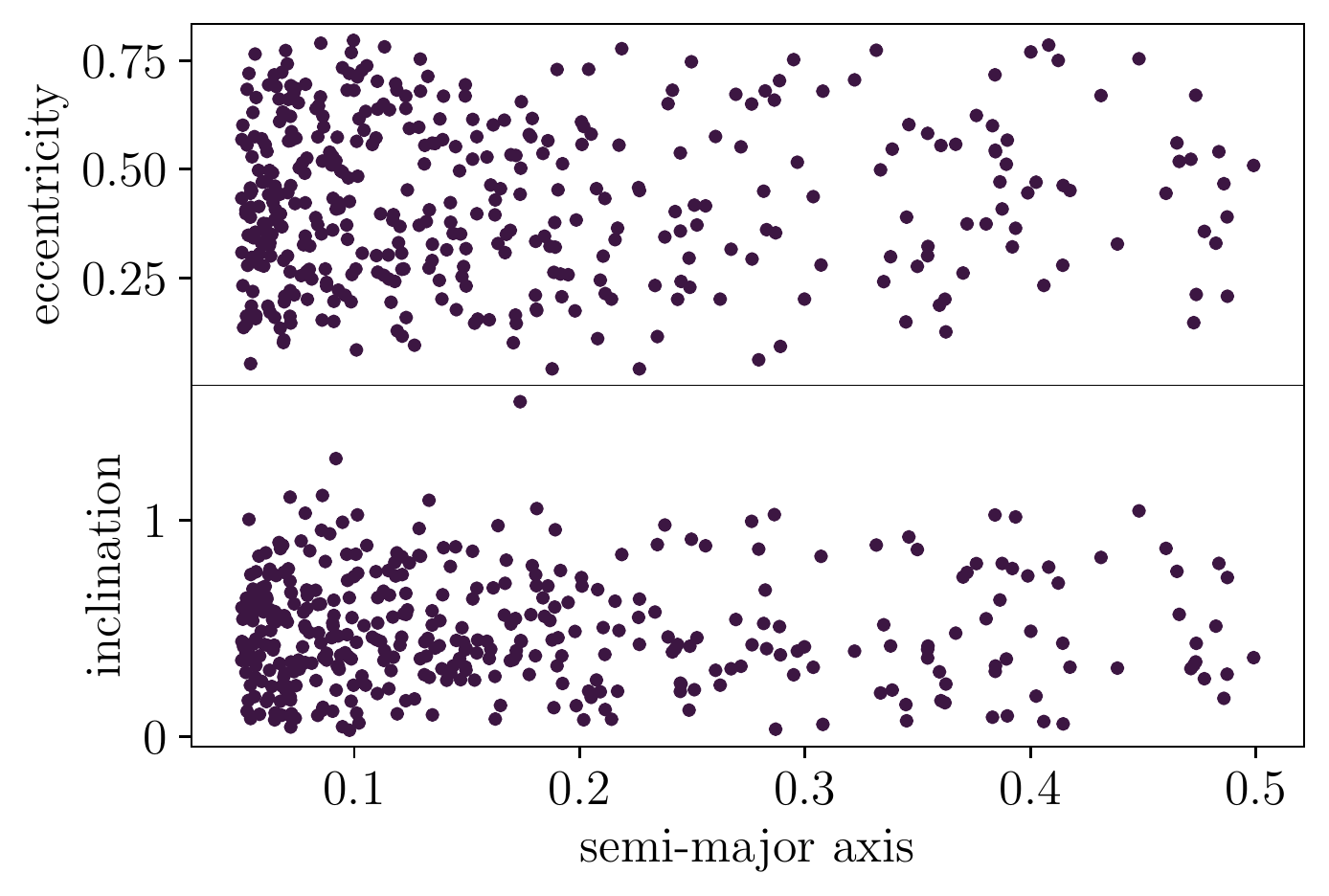}
    \caption{ Initial embryo distributions for the cases with: \textbf{Left:} a surface density slope $\alpha = 1.5$, e$_{avg}$ = 0.1, and inc$_{avg}$ = 0.2;  \textbf{Right:} surface density slope $\alpha = 2.5$, e$_{avg}$ = 0.5 and inc$_{avg}$ = 0.5.}
    \label{fig:init}
\end{figure*}

As mentioned in Section \ref{sec:intro}, we start these simulations with small embryos in an excited disk to maximize the CMF of the planets formed. To get small embryos, we start our simulations at an earlier stage of planet formation than in \citet{2020Scora}, in the runaway growth period when embryos are still forming. To ensure the best conditions for high-velocity collisions and obtain an upper limit to CMFs, we do not include a gas disk, even though it is expected to be present for the first few million years of the formation period \citep{Alexander}. A gas disk should damp the eccentricities and inclinations of the embryos, removing the opportunity for them to experience high-velocity collisions. 

To generate excited initial embryo distributions, we draw the embryo's initial eccentricities and inclinations from Rayleigh distributions. To see how much excitation is needed and emulate the diversity in systems, we vary the mean value of the Rayleigh distributions across three different average eccentricities and two average inclinations (see Table \ref{tab:init-params}).  We impose a maximum eccentricity of 0.8 to ensure that the embryos remain bound. 

The other parameters varied in our simulation are the  surface density slope ($\alpha$) of the population of initial embryos and the mass fraction of debris that is lost due to collisional grinding, referred to as the debris loss. As in \citet{2020Scora}, we pick a maximum mass loss fraction of $50\%$ based on \cite{2012Jackson}, and compare the results to those of a run without debris loss to see the effect of this parameter. Table \ref{tab:init-params} provides the values of each parameter that we use. We run each combination of parameters $\sim$ 3 times (except for those with surface density slopes of $\alpha = 1.5$, that we run only once, as those simulations are significantly computationally slower than the others). We loosely base our embryo mass distribution off of the truncated annulus of \citet{Hansen2009} because it is a simple set-up, ideal for this study's more theoretical approach. As well, it is often used in solar system simulations to emulate the truncation of the disk due to the outer giant planets. To adapt this setup for super-Earth formation, the disk mass is increased to $20 M_{\oplus}$, and the annulus bounds are changed to 0.05 AU and 0.5 AU. We also changed the embryo mass distribution from a flat profile to a sloped embryo mass surface density profile, modelled by the following equation:
\begin{equation} \label{eq:sigma}
    \Sigma=\Sigma_{o}\left(\frac{a}{1 A U}\right)^{-\alpha}
\end{equation}
The disk mass is divided up into 400 embryos of 0.05 $M_{\oplus}$ each. Figure \ref{fig:init} shows two example initial embryo  configurations.

\subsection{Composition}\label{sec:comp}

The planets' CMFs are calculated in post-processing, following the method of \citet{2020Scora}. We assume the embryos are differentiated at the start of our simulations, based on current work that suggests that small bodies can differentiate in a few million years or less \citep{202Carry,2014Neumann,2019Lichtenberg,1992Tonks}. Additionally, we assume that the embryos differentiate between collisions, as the timescale for the cores to merge is orders of magnitude shorter than the timescale between collisions \citep{DAHL2010}. If embryos do not fully differentiate, their CMFs would not change as much with each collision, so this assumption serves the goal of achieving the maximum CMF variation possible. 

Each embryo begins with a CMF of 0.33, based on the iron to silica ratios in our Sun \citep{2014Palme}. The code tracks the embryos and changes their CMF after each collision.  The specific rules governing the change in CMF of the embryos depend on the type of collision and the initial CMFs of the colliding embryos.  Figure \ref{fig:coll-graphic} shows the basics of how core and mantle material are distributed for each collision type. Catastrophic disruption and erosive collisions are the most disruptive collisions that still leave behind an embryo, so they have the greatest potential to increase a planet's CMF. They can strip the largest mantle mass off an embryo, leaving behind a body that is almost entirely core material. Less energetic collisions such as graze'n'merge and perfect mergers result in embryos with the average of the CMFs they started with. In general, core and mantle masses are conserved when calculating the new composition. The new embryo (or embryos) are assigned their new CMFs, and then the remaining material is placed into the debris. Core and mantle mass ratios are not conserved for runs with debris loss, as in these runs some debris mass is discarded and that mass is typically mantle material. More details can be found in the Appendix B of \citet{2020Scora}. 

\section{Final planet systems} \label{sec:system}
\subsection{Average system parameters }\label{subsec:sys-params}
We simulate the formation of 39 systems for a few million years until the planets are each about 10 Hill radii apart, which we consider to be indicative of a stable configuration. The systems could still experience some instabilities later on, but the planets have for the most part formed to their final masses and compositions. Some embryos at the outer edges of these disks are still forming, and retain masses similar to the initial embryo mass. Thus, for this study, we only consider a planet to have formed when its mass is above 0.1 $M_{\oplus}$, twice the initial embryo mass.

\begin{table}
    \caption{System parameters resulting from our simulations with and without debris loss.}
    \centering
    \begin{tabular}{ l | c | c  }
         parameter & debris loss & no debris loss \\
         \hline
         average planet eccentricity & 0.048 & 0.041\\
         average planet inclination & 3.0 & 1.9 \\
         number of planets & 10 & 9 \\ 
         average planet mass & 0.69 & 2.2 \\
         average number of collisions & 11 & 22 \\
    \end{tabular}
    \label{tab:final-params}
\end{table}

Large masses of debris are created throughout the simulations, significantly inflating the computation time. In debris loss runs this results in the loss of about half of the initial total disk mass over the course of the simulations. Thus, the planets formed in these runs remain small, covering a mass range from 0.1 to 2 $M_{\oplus}$, with an average mass of 1.4 $M_{\oplus}$. Planets in the runs without debris loss formed larger, up to 6 $M_{\oplus}$, so the planets simulated in this study range from sub-Earths to super-Earths. The debris loss simulations resulted in more planets per system because of their lower final system masses, consistent with past studies \citep{2014Raymond}. Planets formed with an average of 11 collisions in debris loss runs, half of the number of collisions for planets in non-debris loss runs, due to the loss in total mass (see Table \ref{tab:final-params} to see a summary of the final run parameters). At the extreme end, a few planets experienced more than 50 collisions in the course of their formation. This almost entirely happens in non-debris loss simulations where the initial embryo population began with the flatter $\alpha = 1.5$ surface density slope. Non-debris loss form larger planets with a lower multiplicity than debris-loss runs, so these larger planets form from more embryos. The $\alpha = 1.5$ embryo distribution also has a higher density of embryos to collide with each other in the middle of the embryo distribution than the $\alpha = 2.5$ distribution. Finally, more mixing occurs in planets in the middle of the disk, where all the embryos' orbits overlap. All of this results in embryos from all over the disk colliding with each other to form a planet with a prolific collision history. 

\subsection{Eccentricity and inclination}\label{subsec:ecc}
\begin{figure*}
    \centering
    \includegraphics[width=0.8\textwidth]{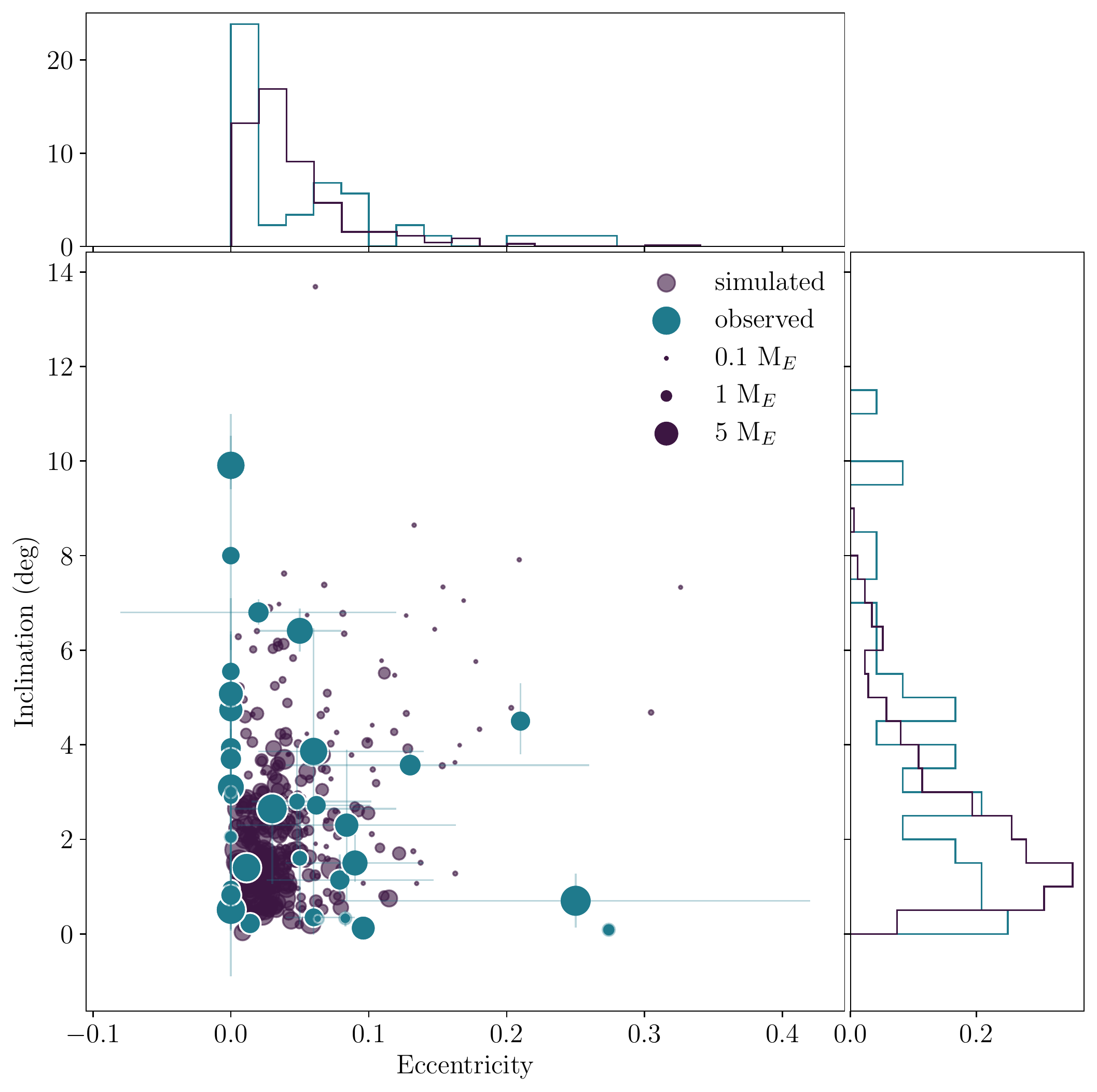}
    \caption{Eccentricities of the planets formed (purple) compared to the eccentricities of super-Earths with well-constrained values from the NASA Exoplanet Archive (teal) \citep{NEA}. The distribution of planets across each axis is shown in the histograms. The size of the dots corresponds to the mass of the embryos. The embryos begin with high eccentricities, but the dynamical friction caused by the debris from collisions brings the eccentricities of most of the final planets down to values comparable to observed exoplanet systems.}
    \label{fig:compare-to-obs}
\end{figure*}

Due to the dynamical friction from the debris and the lack of an external source of excitement to the disk, the average eccentricity and inclination of the embryos drops rapidly over 1 Myr. While the initial embryo eccentricities range from 0-0.8, the final planets fall into a similar range of eccentricities as the observed super-Earths, as seen in Figure \ref{fig:compare-to-obs}. The simulated planets have an average eccentricity of 0.048 for the $50\%$ debris loss runs and 0.041 for non-debris loss runs. The average inclinations were $2.9^{\circ}$ for debris loss and $1.9^{\circ}$ for non-debris loss runs. 

This demonstrates the significant role that debris can play in damping the orbits of embryos as they form. The high mass of debris completely erased the high-eccentricity origins of the majority of the simulated planets. Thus, it seems that an excited disk origin cannot be ruled out from present-day observed low eccentricities and inclinations.  Other factors, like the number of planets in the system, may point towards the initial conditions of the disk.

\subsection{Occurrence of high energy collisions }
\begin{figure*}
    \centering
    \includegraphics[width=0.9\textwidth]{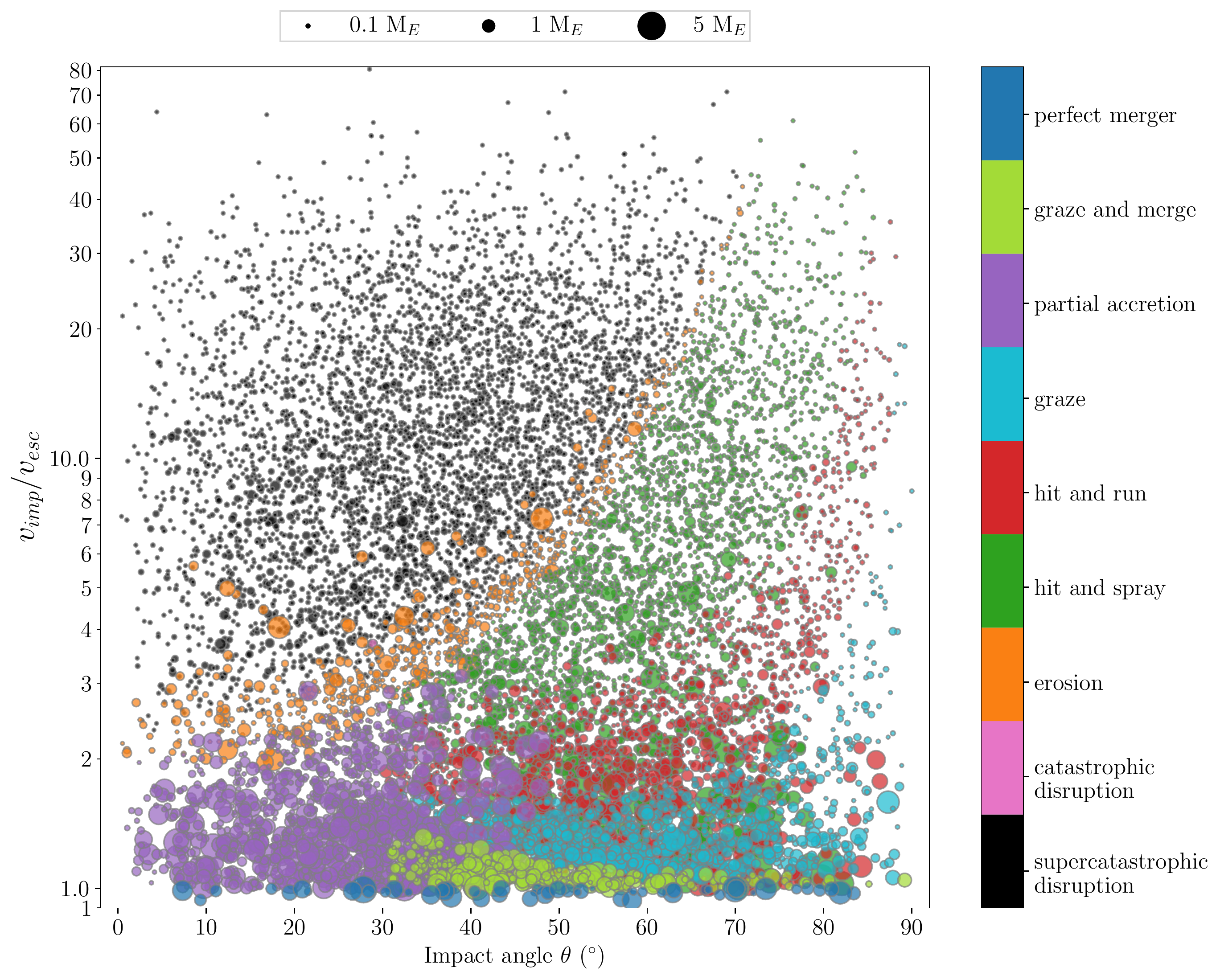}
    \caption{Impact angle and impact to escape velocity ratio of all the collisions in the simulated runs. The size of the points corresponds to the total colliding mass of the collision. The collisions in the simulations cover the entire parameter space of collision velocities and angles. Early on in the formation process, when embryos are small, there are large numbers of super-catastrophic collisions. The distribution of collisions for larger embryos later on in the formation process sticks to lower velocities and other, less destructive collision types.}
    \label{fig:impact-params}
\end{figure*}

\begin{figure}
    \centering
    \includegraphics[width=0.45\textwidth]{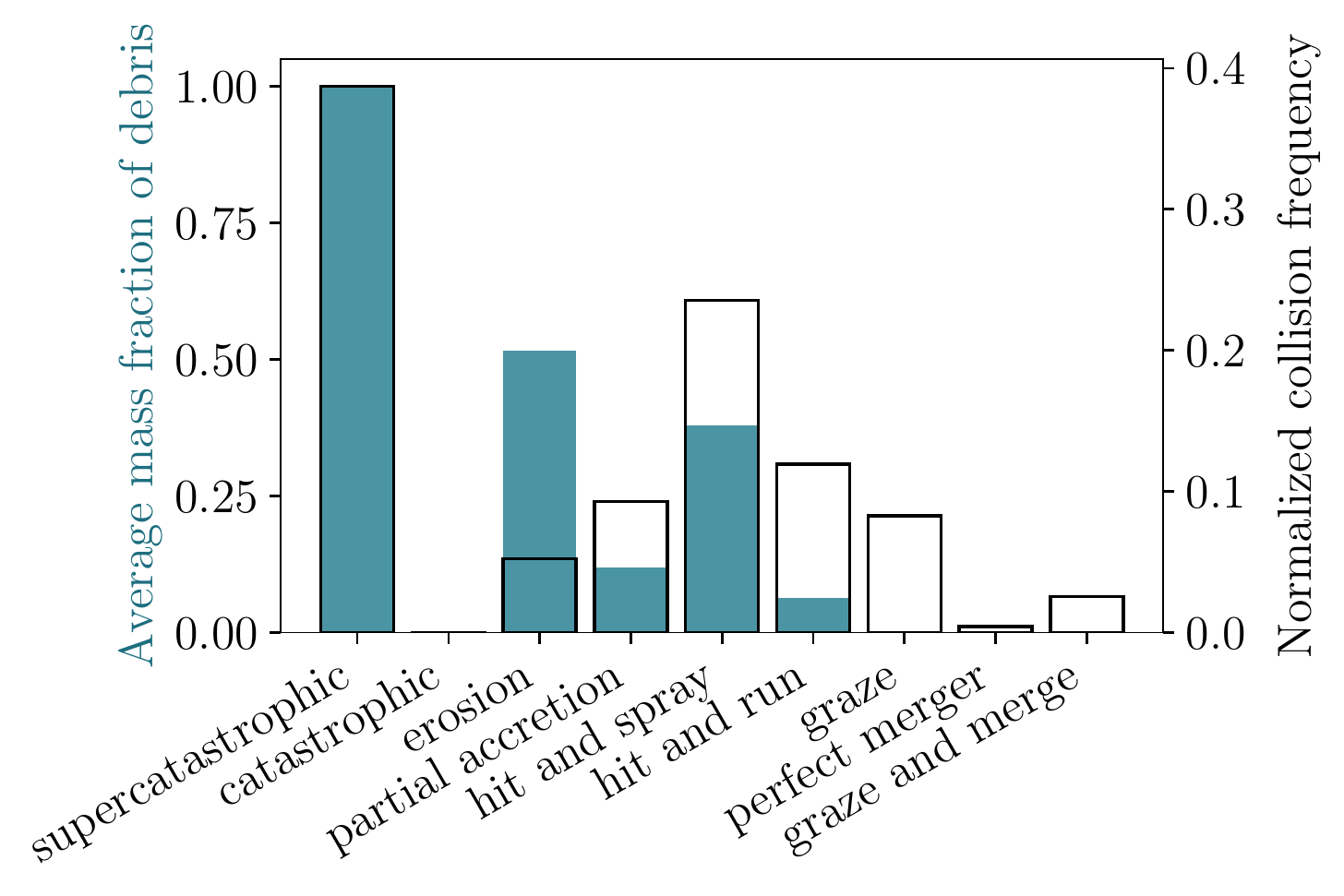}
    \caption{The frequency of each type of collision (black outline), compared to the average mass of debris that the type of collision creates (blue). Super-catastrophic collisions are the most common and create the most debris. The second most common type of collision is the hit'n'spray. Erosive collisions create a significant amount of debris and can modify the CMF of the remnant, but make up only 8\% of the collisions that occur.}
    \label{fig:debris-prod}
\end{figure}

Figure \ref{fig:impact-params} demonstrates that the high excitation of the disks successfully produced many high-velocity collisions. In \citet{2020Scora}, no collisions happened above an impact velocity to mutual escape velocity ratio ($v_{imp}/v_{esc}$) of 8 due to both a dynamically cold disk and large gravitational wells resulting from the large masses involved in the collisions. In simulations starting with excited disks and lower embryo masses, the maximum $v_{imp}$ ratio is an order of magnitude larger.  Super-catastrophic disruption collisions, which start happening above $v_{imp}$ ratios of  3, were the most common collision type (see Figure \ref{fig:debris-prod}). As mentioned above in Section \ref{subsec:ecc}, the eccentricity and inclination of the embryos decreased over time due to dynamical friction, so super-catastrophic disruption collisions and other high-velocity collisions occur almost exclusively early on in collisions between small embryos. As the evolution proceeds and the embryos become more massive, they mostly experience collisions with lower $v_{imp}$ ratios.

Erosive collisions are not as destructive as super-catastrophic disruption collisions, however, they produce the most CMF increase in our simulations. This is because no embryo remains after super-catastrophic disruption collisions, thus their main effect is in producing debris that later accretes onto other planets. Erosive collisions are the third-most destructive collisions (catastrophic, being the second-most destructive, do not occur in any of our runs). Each erosive collision strips a significant amount of mass from the target embryo, but still leaves behind an embryo that can go on to grow into a planet. On average they turn $\sim 55\%$ of their total colliding mass into debris in our simulations (see Figure \ref{fig:debris-prod}). Consequently, these collisions are capable of removing the entirety of an embryo's mantle, often creating embryos with a CMF of 1. Compared to \citet{2020Scora}, erosive collisions are $\sim 5$ times more frequent in our excited disk simulations.

\subsection{Final planet compositions}\label{subsec:final-results}
\begin{figure*}
    \centering
    \includegraphics[width=0.8\textwidth]{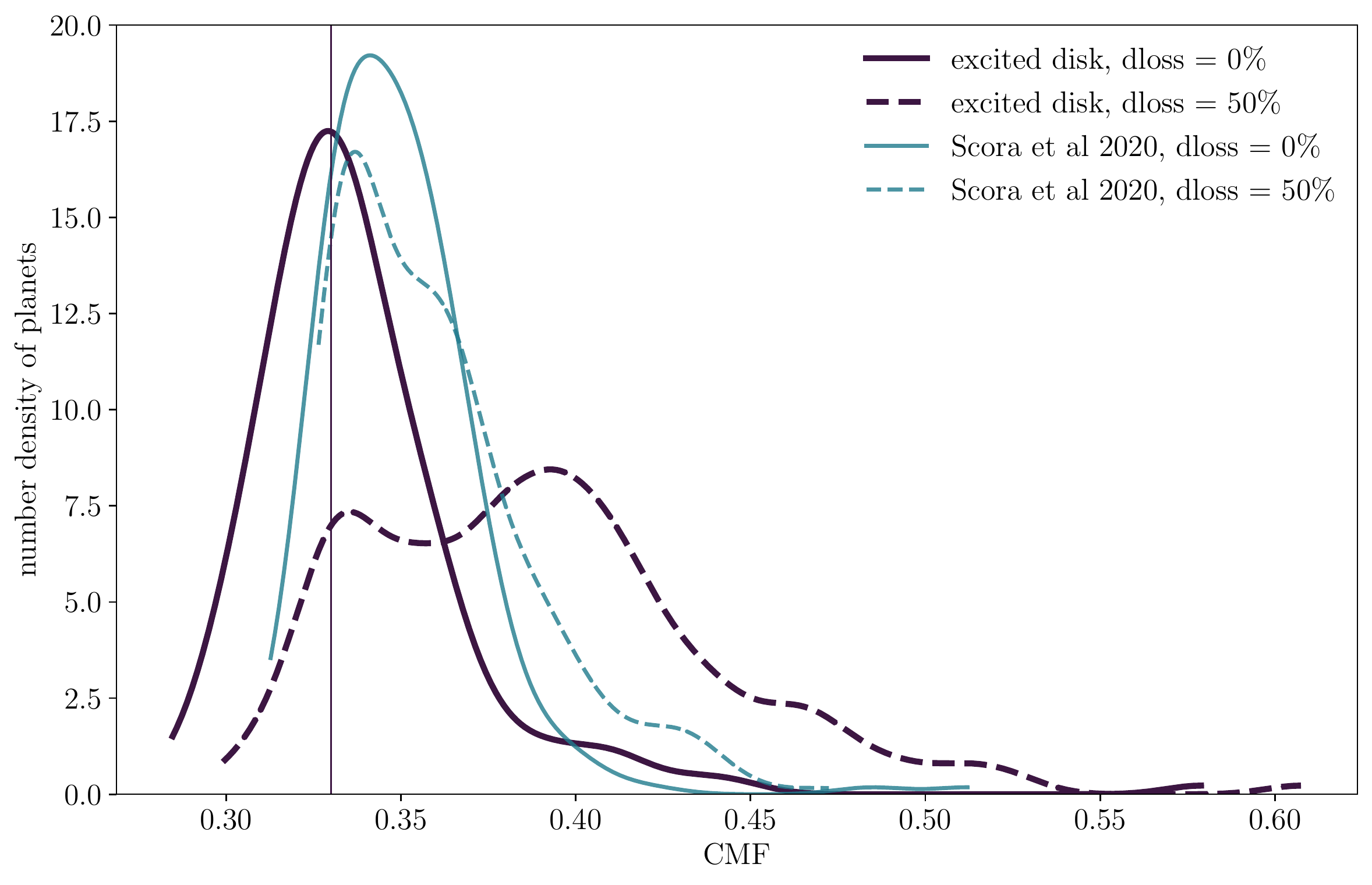}
    \caption{Comparison of the CMF distributions of the final planets in debris-loss runs to those without for this study and for planets from \citet{2020Scora}. The distributions are achieved with a kernel density estimator (kde) to avoid binning effects. The excited disk simulations with $50\%$ debris loss have a significantly wider distribution than the others, and therefore a higher average CMF. The maximimum CMF reached is higher for planets in this study than \citet{2020Scora}, but the maximum CMF is roughly the same for both debris loss and non-debris loss runs. }
    \label{fig:CMF-dist}
\end{figure*}

With the increase in high velocity collisions comes a significant increase in the number of planets with CMFs between 0.4 and 0.5 compared to \citet{2020Scora}, as seen in Figure \ref{fig:CMF-dist}. Some planets have CMFs of almost 0.6, approaching the lower end of Mercury's CMF range (0.7) \citep{2013Hauck}, yet these planets are still exceedingly rare. Thus, while we have succeeded in forming more high CMF planets, Mercury-like planets remain hard to form in our simulations.

We use a kernel density estimator (kde) to smooth out our CMF distributions and avoid histogram binning effects. The CMF distribution of our final planets in Figure \ref{fig:CMF-dist} appears Gaussian for planets with no debris loss, peaking at the initial CMF given to the embryos and with a small tail of high CMF planets. The average of this distribution is a little higher than the initial CMF at 0.34. This contrasts significantly with the distribution of planets with $50\%$ debris-loss. While there appear to be multiple peaks in this distribution, this is just due to the small sample size of planets. The overall shape of the distribution is wider and no longer Gaussian, and therefore the average CMF of the planets produced in debris-loss simulations shifts up to 0.39.  This demonstrates that debris loss plays a much more significant role in shaping the CMF of the planet population in our excited disks than it did in the simulations of \citet{2020Scora}, shown in Figure \ref{fig:CMF-dist} as the teal curves. Section \ref{sec:CMF-mass-coll} goes into detail as to why the $50\%$ debris loss planets have such a substantially different CMF distribution. 

In addition, there is a clear difference in the location of the CMF distribution peaks of the planets with no debris loss in this study and the simulations with low eccentricities in \citet{2020Scora}. The peaks of the CMF distribution in the previous study is shifted by $\sim 0.02$ towards a CMF slightly higher than the initial, while for this study the peak of the distribution of planets without debris loss is centered on the initial CMF. This centering of the CMF distribution can be explained by the increased mass of debris generated in this study when compared to the previous work. The phenomenon is explained in further detail in Section \ref{subsec:debris-reacc}.

Figure \ref{fig:obs-super} shows the distribution of rocky exoplanets with good precision mass and radius measurements. There are significantly fewer planets in this dataset than in our simulated datasets, and this has not been corrected for observational bias. For the purposes of this theoretical study, however, we make some general comparisons between the distribution of observed CMFs and those of our simulated planet CMFs. The planets are well spread out across the CMF range, such that they better match the distribution of planets with $50\%$ debris loss than those without. Additionally, the maximum CMF of the observed planets is around Mercury's CMF of 0.7 \citep{2013Hauck}, thus the maximum CMF of our simulated planets is still too low to reproduce the compositional variety observed in super-Earths. However, if we consider the error bars on the data points, the maximum CMF is not as clear, and may fit better with our observations. Thus, we need more accurate measurements of mass and radius to better compare our results to observational data.

\section{Relationship between CMF and collision history}\label{sec:CMF-collhist}
\begin{figure*}
    \centering
    \includegraphics[width=\textwidth]{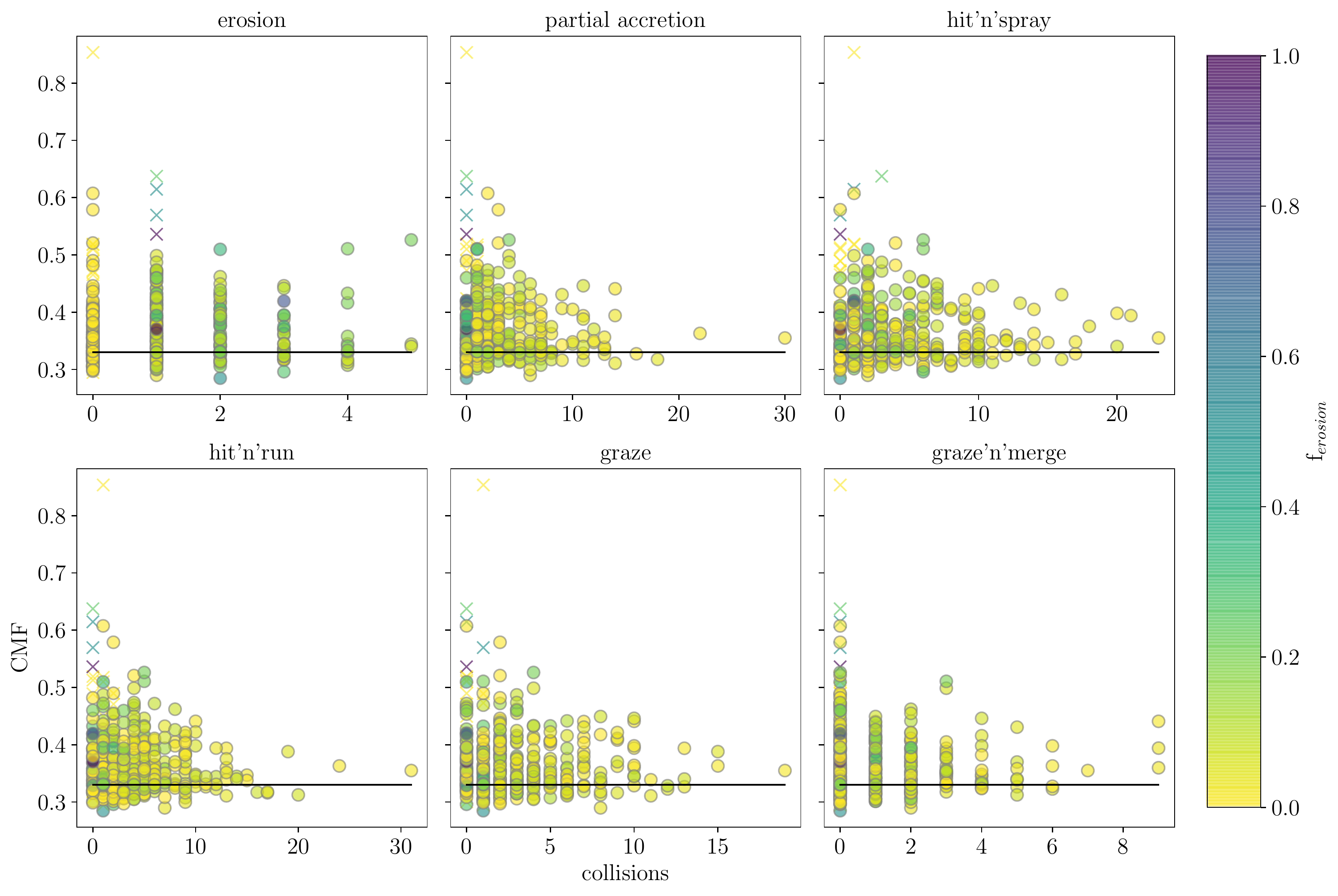}
    \caption{Final CMF of each planet with respect to the number of each type of collision it experienced in its lifetime. The planets are colour-coded by $f_{erosive}$, which is the fraction of the collisions in the planet's history that were erosive collisions. Planets less than 0.1 $M_{\oplus}$ are included as 'x' in the plot. The black line is the initial CMF all embryos start with. Super-catastrophic collisions are not included because any embryo that experiences these will not become a planet. We exclude catastrophic disruption because it does not occur, and merge is not included but follows the same trend seen in the panels shown. For all collision types, a planet that experiences more collisions of that type will be more likely to have a final CMF that is closer to the initial CMF. The fewer collisions of a certain type the planet experiences, the wider a range of final CMFs it can occupy. For erosive collisions, this trend is not as strong, particularly since there are so few planets with more than 2 erosive collisions.}
    \label{fig:coll-types-CMF}
\end{figure*}

\begin{table*}
\centering
\caption{A description of the collision types used in our planet formation simulations. See \citet{Leinhardt2012} and \citet{Genda2012} for the quantitative boundaries between these collision types and the fraction of the colliding mass ejected into space, and Figure \ref{fig:coll-graphic} for visual representations of these collisions.}\label{tab:coll}
\begin{tabular}{ l | c | l}
Collision Name & Embryos remaining & Description \\
\hline
Supercatastrophic disruption & 0 & Only debris remains \\
Catastrophic disruption & 1 & Small amount of target remains, mainly debris  \\
Erosion & 1 & The target is eroded by projectile  \\
Partial accretion & 1 & The projectile is partially accreted by the target  \\
Hit and spray & 1 & Grazing collisions where the projectile is eroded \\
Hit and run & 2 & Two bodies graze each other and produce some debris \\
Graze & 2 & Target and projectile bounce inelastically at large angles \\
Perfect Merger & 1 & Target and projectile merge \\
Graze and merge & 1 & Target and projectile impact and then merge \\
\end{tabular}
\end{table*}

\begin{figure*}
    \centering
    \includegraphics[width=0.8\textwidth]{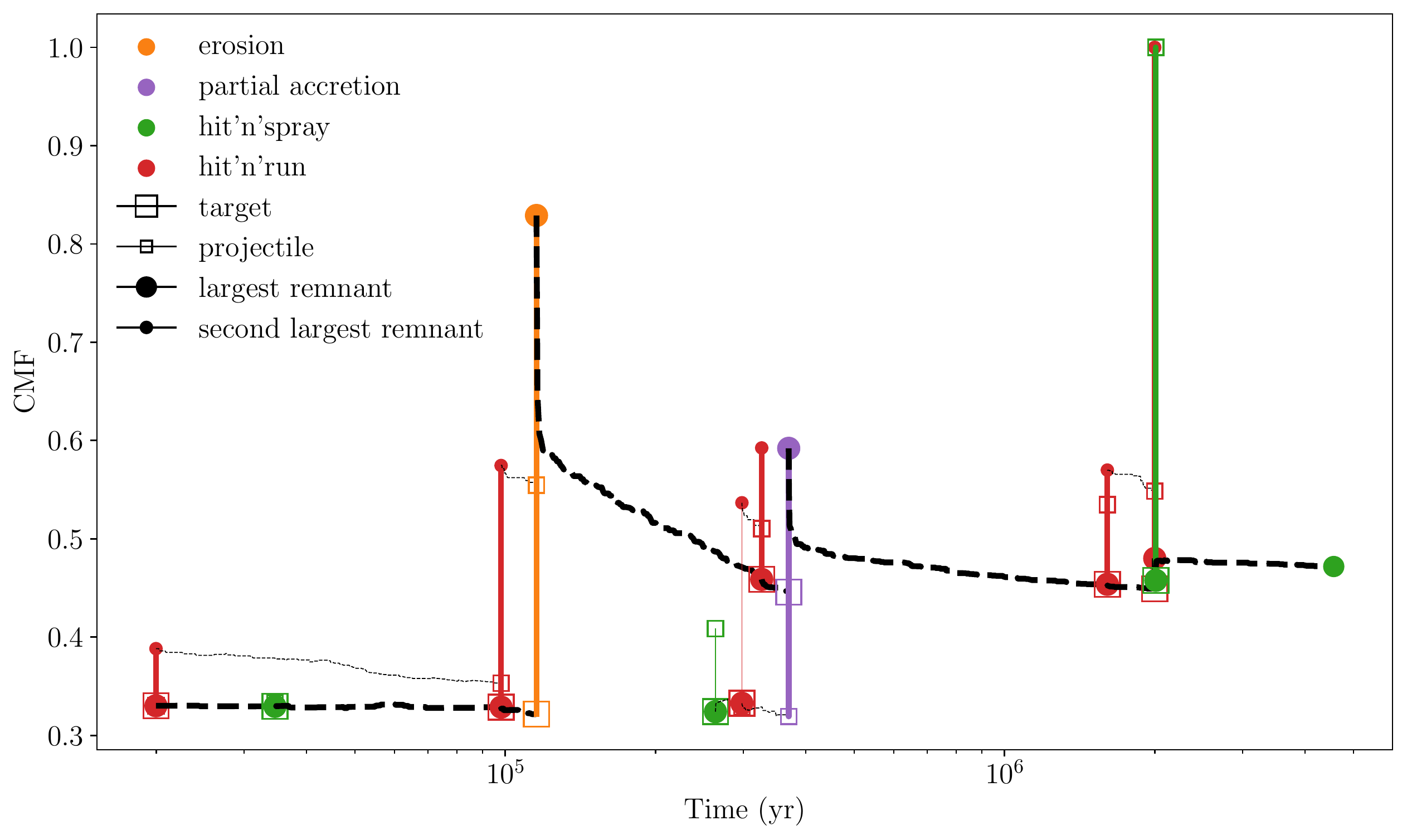}
    \includegraphics[width=0.8\textwidth]{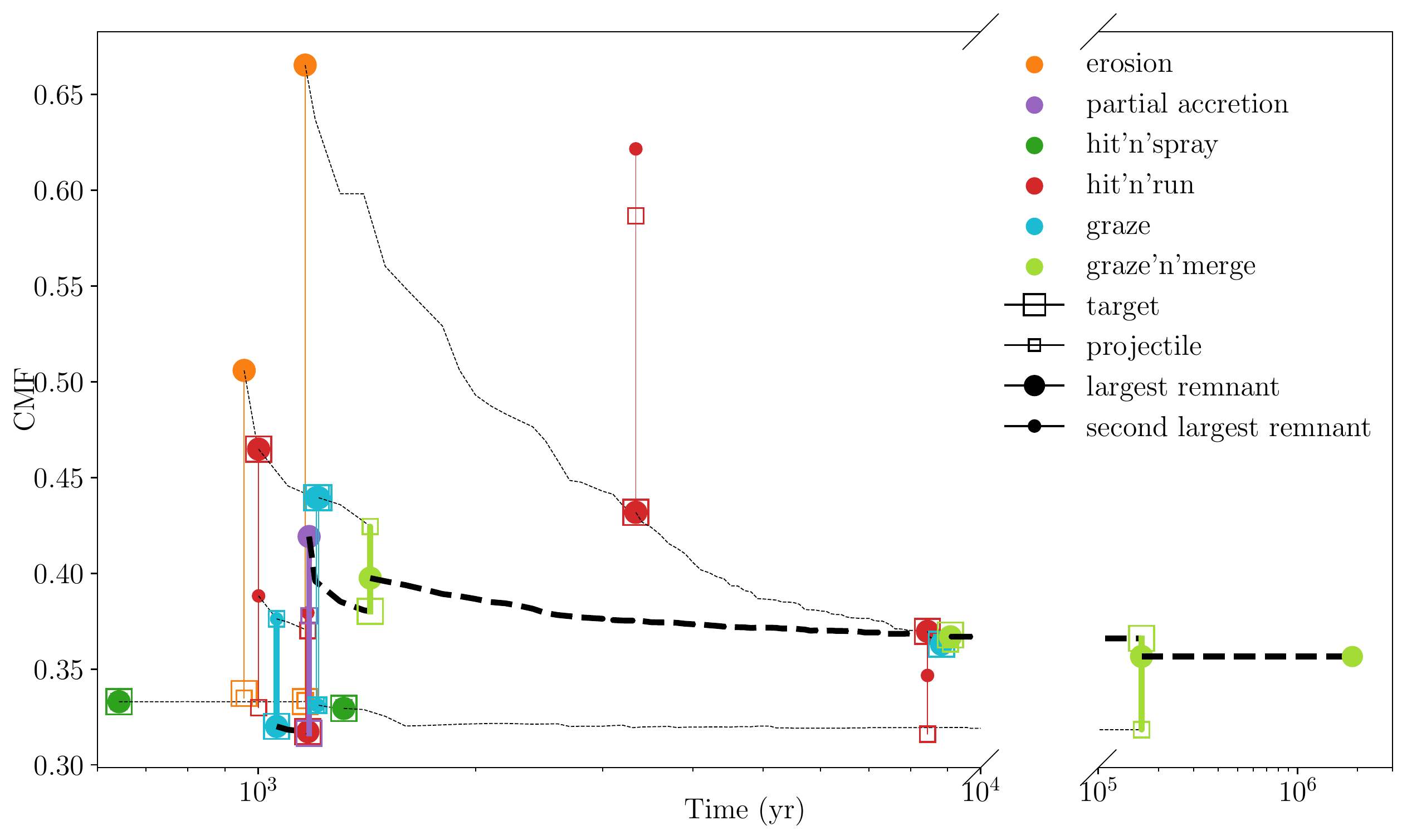}
    \caption{Collision histories of two planets from debris-loss cases. Each collision is represented by a set of coloured dots and lines. Unfilled dots represent the CMF of the bodies that collide, and the filled squares represent the bodies that remain after the collision. The 'largest remnant' is the most massive (and only, for most collisions) embryo that remains. Some collisions have two remaining embryos (see Table \ref{tab:coll}), and in this case the second embryo is called the 'second largest remenant'. All of the dots for one collision are connected by a solid, coloured line. Dashed lines represent the body's CMF over time between giant collisions. CMF change between giant collisions is due to debris accretion. The path of the main embryo that becomes the planet is bolded. \textbf{Top:} The planet has a final CMF of 0.49. It experiences only one erosive collision, but other collisions it experiences close to the end of its life keep the planet's CMF high. (i.e. hit'n'run, hit'n'spray), and debris reaccretion does not bring its CMF down as much. \textbf{Bottom:} Final planet CMF of $\sim 0.35$. Even though it experiences two erosive collisions that significantly increase its CMF, the final CMF of the planet is actually quite close to the initial of 0.33 because of debris reaccretion and other collisions that bring the planet's CMF down, mainly graze'n'merge.}
    \label{fig:collhist-highCMF}
\end{figure*}

The basis of this study was the hypothesis that the more erosive collisions a planet experiences, the higher its final CMF would be; however, the results prove it is not this simple. In Figure \ref{fig:coll-types-CMF}, we see that the more erosive collisions a planet experiences in its lifetime, the lower its final CMF will be (with one exception). In fact, this holds true for all collision types, six of which are shown in Figure \ref{fig:coll-types-CMF}. As planets experience more collisions of any type, the range of final CMFs they occupy becomes smaller and the maximum of this range decreases to approach the initial CMF. This trend stays true for the total number of collisions that a planet experiences as well. There are two main reasons for this trend of decreasing CMF with increasing number of collisions: the rareness of erosive collisions, and debris reaccretion. 

\subsection{Infrequency of erosive collisions}\label{sec:erosion}
The first reason for the trend of decreasing CMF range with increasing collision frequency is that erosive collisions are rare collisions. 
They make up only $ 5\%$ of the total collisions, or $6\%$ of the collisions that form planets.  It is therefore extremely unlikely for a planet to experience only erosive collisions during its formation. Indeed, we see this reflected in Figure \ref{fig:coll-types-CMF}, where no planet has more than 5 erosive collisions, yet planets can experience up to 10 or 20 of the more common collision types. Therefore, when we consider all of the collisions in a planet's history, the fraction of those that are erosive collisions, $f_{erosive}$, is typically quite small. Only two or three planets have a $f_{erosive} > 0.8$.  Thus, a planet with multiple erosive collisions in its history may experience large increases in CMF, but it will also likely experience many more non-erosive collisions that can have a stronger effect overall on its CMF. This can be clearly seen in the fact that this trend of decreasing CMF range with increasing collision frequency holds for graze collisions. These collisions do not change a planet's composition or mass at all, thus, the effect we see here is that high numbers of graze collisions corresponds to high numbers of other types of collisions that are affecting the planet CMF. 

To investigate in detail the reason a planet finishes with a particular compositional make up, we construct its collisional history tree using \texttt{colltree}\footnote{This code is publicly available on Github at \url{https://github.com/jscora/colltree}, and version 0.1.1 is archived in Zenodo \citep{scora_colltree_code}}. It shows which collisions were involved at what time, and how the CMF changes throughout the planet's evolution. Two representative examples are shown in Figure \ref{fig:collhist-highCMF}.  One where the final planet's CMF is high (left), and one where the planet's final CMF is closer to the initial one (right). Both planets experience large increases in CMF over the course of their formation due to erosive or other high-energy collisions. These high-energy collisions mainly happen early in the planet's formation history. However, throughout their histories both of them also experience other collisions that bring the CMF of the planet down, and so the final CMF is determined mainly by the collisions that occur late in their lifetime. The planet with a high CMF (left) experiences a partial accretion that significantly increases its CMF near the end of its formation, while the one with a low CMF (right) has graze'n'merge collisions that decrease its CMF. These two planets illustrate a general trend seen in the formation histories of the simulated planets, where many planets tend to experience collisions that increase their CMF once or even twice in their formation, but their CMF is brought back down after and thus this is not reflected in their final CMFs.    

Clearly, the majority of the non-erosive collisions that make up most of the planet's collision history tend to have an averaging effect on the planet's CMF over time. Typically, partial accretion collisions will increase the CMF of the planet, but this increase is smaller than those caused by erosive collisions. Hit'n'spray and hit'n'run collisions can cause either increases or decreases in the final planet's CMF, depending on the details of the collision. Graze'n'merge and merging collisions average the CMF of the planet with the body it collides with, which will tend to decrease the CMF of a high-CMF planet. Therefore, all of these collisions tend to balance each other out, bringing the planet's CMF closer to the initial value of 0.33 and counteracting the effects of the few erosive collisions it may have.

\subsection{Debris reaccretion}\label{subsec:debris-reacc}
The second averaging force on planet CMFs is debris reaccretion. Planets will often reaccrete the debris that forms from a collision, as has been shown in other simulations with imperfect collisions \citep{Chambers}. We can observe this effect in Figure \ref{fig:collhist-highCMF} in the curved, dashed lines after a collision that has increased a planet's CMF. These lines show the change in the planet's CMF over time due to debris accretion. Debris material from collisions is almost always mantle material, hence the downward slope of these dashed lines. Due to the high number of other embryos in the disk, only part of the mantle debris will reaccrete onto the embryo, leaving an embryo that does not have as high a CMF as it did, but one that is still enriched compared to before the collision.  In runs with $50\%$ debris mass loss, there is even less mantle debris remaining to be reaccreted, leaving embryos with even higher CMFs after the collision. This is the main reason that the CMF distribution of planets formed with debris-loss is so different from that of those formed without. Thus, while this has an averaging effect on the CMF of planets, it does not completely erase CMF changes that are caused by collisions.

Debris accretion has a much stronger averaging effect in simulations with no debris loss, and this is what contributes to the shift in the peak of the planet CMF distribution seen in Figure \ref{fig:CMF-dist}. As mentioned in Section \ref{subsec:final-results}, the peak of the CMF distribution for simulations with no debris loss is centered on the initial CMF, whereas for the previous study with lower eccentricities \citet{2020Scora}, the peak is slightly higher than the initial at 0.35 CMF. This is due mainly to the increase in debris mass in the eccentric simulations. In the simulations in this study, there are more erosive and high-velocity collisions than in \citet{2020Scora}, and these collisions create higher masses of debris. Thus, debris reaccretion has a stronger effect in these simulations than in \citet{2020Scora}. There is also a significant mass of debris created by super-catastrophic collisions early on in the formation process. These collisions completely destroy an embryo and leave only debris behind, that then spreads out across the orbits of the remaining embryos and is reaccreted. Most of this debris comes from embryos with the initial CMF of 0.33, thus accreting significant amounts of this debris also has a general averaging effect on planet CMFs. Therefore, the increase in debris mass produced in these simulations centers the CMF distribution of planets more strongly around the initial CMF than in \citet{2020Scora}.

\begin{figure*}
    \centering
    \includegraphics[width=0.8\textwidth]{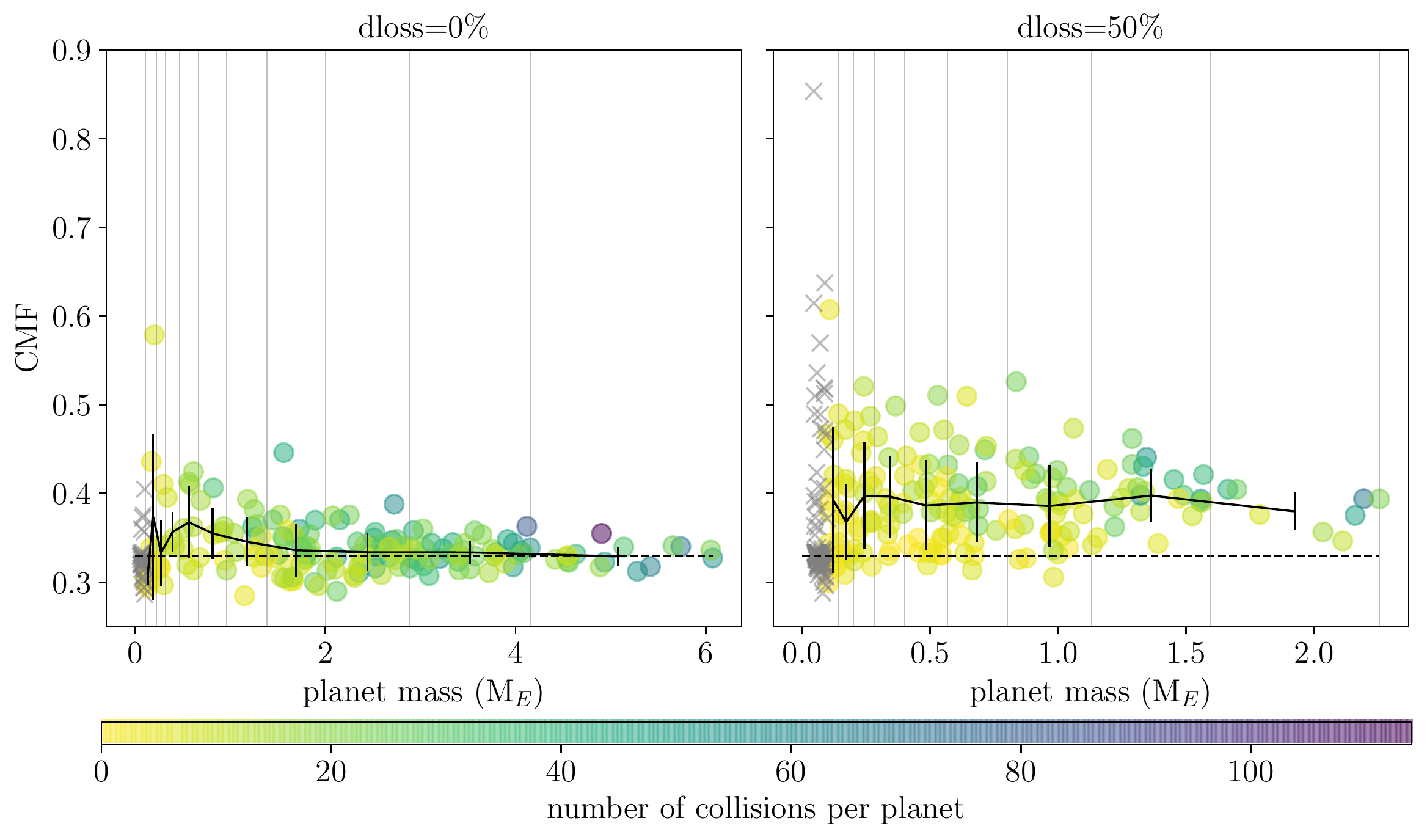}
    \caption{Relationship between the final CMF and mass of planets for runs with $0\%$ and $50\%$ debris loss. The points are colour-coded by the number of giant collisions experienced in their lifetime. The colourbar is on a log scale from 0 to 100. The dashed line shows the initial CMF given to all planets, and the solid black line with error bars shows the mass-binned average CMF. The vertical grey lines show the bins, and the error bars show the $1 \sigma$ spread in CMF in each bin. The grey 'x's are planets that are smaller than 0.1 $M_{\oplus}$. We see a strong correlation between the planet mass and the range of CMFs that it can have, with more massive planets that have experienced more collisions having lower CMFs.}
    \label{fig:mass-CMF}
\end{figure*}

\section{Relationship between CMF and planet mass}\label{sec:CMF-mass-coll}

Both the averaging effect of non-erosive collisions and debris accretion combine to generate the relationship between mass and CMF that we see in Figure \ref{fig:mass-CMF}. At low masses, planets are spread over a wide range of CMFs. The more mass a planet accretes, either from debris reaccretion or mass-accreting collisions such as graze'n'merge, the closer its CMF approaches some average. This is due to the fact that energetic disruption collisions that increase a planet's CMF all decrease the target planet's mass. The collisions that increase a planet's mass are all types where some or all of the projectile is accreted (i.e. partial accretion, hit'n'spray, perfect merger). In most cases, the core of the projectile is accreted first, so this can cause an increase in the planet's CMF. However, overall this results in a smaller increase in CMF than removing mantle material, and typically there is both core and mantle material being accreted, which further reduces the ability of the collision to increase the planet's CMF. As discussed in Section \ref{sec:erosion}, therefore, these collision types tend to average out the CMF of the planet. On top of this, the debris mass that is accreted is typically mantle material, which lowers the CMF, as explained in Section \ref{subsec:debris-reacc}. Thus, planets that have accreted the most mass have compositions that have experienced the most averaging effects of giant collisions, as well as the most CMF decrease due to debris reaccretion. The resulting relationship between CMF and mass looks very similar to the relationship between CMF and collisions in Figure \ref{fig:coll-types-CMF}, and the increase in planet mass is generally accompanied by an increase in the number of collisions it experiences. The mass of a planet and the number of collisions it has experienced are not one-to-one, but they are strongly related.

Including $50\%$ debris mass loss in these simulations results in an increase in the average CMF that the planets approach as they increase in mass, and a spreading out of the planet CMF distribution. In runs with no debris loss, the average is very close to the initial CMF of 0.33. In the systems with debris loss, the average appears to be slightly higher, around 0.4 CMF. This increase in average CMF for the debris loss planets is due to the changes to debris reaccretion caused by the $50\%$ debris mass lost per collision. Since the debris mass created is typically mantle material, the debris mass lost is necessarily mostly mantle material. This leaves behind less mantle material to be reaccreted by planets. Thus, planets do not gain as much mass, and when they do the CMF-lowering effect is less strong. Over time, the loss of mantle debris will therefore increase the average CMF of the material in the simulation, resulting in a similar increase in average CMF of the planets. 

Losing $50\%$ of the debris mass after each collision has a non-linear effect, reducing the overall masses of the planets by a factor of 3, as discussed in Section \ref{subsec:sys-params}. The lower mass range of the population of planets in the debris loss runs could be a confounding variable contributing to its higher average planet CMF. However, the binned average for the runs with debris loss in Figure \ref{fig:mass-CMF} remains higher than that of those without debris loss across the mass range. Thus, the higher average CMF of the debris loss runs is not simply due to their different mass ranges. The effect of debris loss on the average planet CMF is the cause of the wider CMF distribution seen in Figure \ref{fig:CMF-dist}. The averaging effect of debris reaccretion is less strong, and thus the peak around the initial CMF is much weaker, and the distribution skews towards higher CMFs.

\section{Implications}\label{sec:implications}

\subsection{Caveats}
The initial conditions used in this study are chosen to maximize the impact velocities of the embryos, but they are not in line with the most up-to-date theories of super-Earth planet formation. In particular, they start with small embryos ($0.05 M_{\oplus}$) and no gas disk. It is highly unlikely that there would be no gas disk so early in the formation process when embryos are this small \citep{Alexander}. Any gas disk would damp the collisional speeds of the embryos, resulting in fewer high-energy collisions. However, in the presence of eccentric giant planets the planetary embryos may have an equilibrium non-zero eccentricity even with the gas disk present. In the case that the giant planets are not eccentric but are migrating, the embryos trapped in resonances should also acquire equilibrium non-zero eccentricities. In addition, it is possible that so early on in their formation the embryos would not yet be fully differentiated, thus collisional stripping would not preferentially strip mantle material. In this case, the collisions the embryos experience should provide enough heating to trigger differentiation, and the embryos would be differentiated early on in the simulations. Therefore, the outcomes of our simulations should be considered an upper limit on the CMF change that can be caused by collisions in a scenario where the embryos are colliding from early on. 

Due to the long simulation time required for these runs, there are not many simulations for each parameter set. Thus, while there are enough simulations to draw conclusions about the overall distributions of CMF and collisions, there are too few planets and too few simulations of each parameter set to draw conclusions about smaller effects or outliers. For example, the bumps in the CMF distribution of planets with debris loss are too small to be significant considering the number of planets. 

It is important to note that  $\sim 10\%$ of planets in our simulations that experience only one or two collisions in their lifetime are still unstable. The majority of these low-mass, low-collision embryos have larger semi-major axes and so have not experienced as many orbits as the others. As such, they are likely to continue their evolution as the system evolves further. Therefore, we have chosen not to use planets below $0.1 M_{\oplus}$ in our main dataset. The results may change slightly in the low-mass region in the final stages of planet formation, and including these lower-mass planets could possibly make some minor changes to the overall distributions.

\subsection{H number}
\citet{2021Cambioni} simulated the formation of the solar system and found that planets with CMFs greater than 0.4 typically  experienced a higher number of hit'n'run collisions than their lower CMF counterparts. \citet{2021Cambioni} quantified the 'h-number', which  essentially tracks the number of times a planet has been the 'runner', or second largest remnant, in a hit'n'run collision. This is based on the fact that the second largest remnant tends to have the most mantle stripped from it, thereby undergoing the most CMF change. We tested this result against our own simulations, but we found that our planets have smaller h-numbers, and we did not find a strong correlation between h-number and high CMF. It is possible that this is due to the differences in initial conditions between our simulations. For example, our high initial eccentricities result in more high-energy erosive collisions that dominate the CMF change. As well, we track debris formation in our simulations. They theorized that the debris would typically be reaccreted by the largest remnant, thus their results would not be affected by the presence of debris. In our simulations this hypothesis holds for a large fraction of the 'runners', but there are some cases where a significant portion of debris reaccretes onto the 'runner'. Additionally, the runners may accrete remaining debris from supercatastrophic collisions, and in both cases this debris accretion results in an averaging out of the final planet's CMF.

\subsection{Generating high CMF planets}
Based on the results of this work and \citet{2020Scora}, there are three possible pathways to generate a high CMF planet with erosive collisions. The first is to begin with highly eccentric embryos and remove all debris mass created. This would result in a significant number of high CMF planets, but based on the simulations presented in this paper this seems to be an unlikely formation pathway. We start with 20 $M_{\oplus}$, and planets only form to $\sim 2 M_{\oplus}$. The planets with higher CMFs of around 0.5 are all less than $\sim 1.5 M_{\oplus}$, and those with truly high CMF (i.e. 0.6 or 0.7 CMF) are less than 0.5 $M_{\oplus}$. Extrapolating from this, an excessive amount of starting mass would be needed in the disk to form high CMF super-Earths of around 4 $M_{\oplus}$ such as Kepler 105c \citep{2016Jontof-Hutter,2017Hadden}.

The other method to form a high mass, high CMF planet would be to have a few high energy collisions at the end of formation. This could be achieved by two planets, almost fully formed, with high mutual eccentricities or inclinations. This method would require specific circumstances to produce such a scenario, as collisions are highly stochastic. For example, when two planets are caught in the same resonance with a third, more massive, migrating planet, they will have high eccentricity orbits and a large probability to collide. Even this would not ensure a high CMF planet, but only increase the odds of such an occurrence. 

Finally, another possible pathway is the presence of giant planets perturbing the embryos throughout their formation. Our simulations do not include the presence of giant planets as a first step, but only an instability arising from possible interactions between them. Including giant planets should continuously excite the embryos, counteracting the dynamical friction from the debris. Thus, embryos maintain higher eccentricities throughout formation, resulting in more high-energy collisions occurring over the planet's entire formation history. Consequently, this might weaken the averaging effect of collisions seen in this study, increasing the likelihood of forming more massive, high CMF planets. As with the first scenario, this would require a more massive initial disk due to the increased debris creation that comes with an increase in destructive collisions. 

While each of the above is possible, they require specific circumstances and it is unclear by how much they would increase the prevalence of high CMF planets. Therefore, it seems likely that high mass, high CMF planets are formed by some combination of processes. For example, \citet{2021Adibekyan} found that many high CMF super-Earths orbit stars enriched in iron compared to our Sun, but not enough to generate planets with such high CMFs alone. Another possibility is that iron-rich planetesimals form close to their star due to the streaming instability, forming planets that are already iron-rich before undergoing the giant collisions phase \citep{2020Aguichine,2022Johansen}. Either of these methods would provide iron-enriched embryos that would only require minimal mantle stripping in order to form high-CMF planets.

\section{Discussion}\label{sec:disc}

While we replicate the magnitudes of observed super-Earths eccentricities in our simulated systems, our simulated planets do not have the same distribution of eccentricities. Observed super-Earths tend to fall into one of two categories: low-eccentricity systems of planets, or high-eccentricity single planets \citep{2016Xie,2019VanEylen,2019Mills}. This bimodal distribution is visible in the eccentricity histogram of observed super-Earths in Figure \ref{fig:compare-to-obs}. The simulated planets follow a smoother distribution from low to high eccentricity planets. This can be attributed to the lack of giant planets throughout our simulations, as we only invoke their presence for increasing the inclination and eccentricities at a given time. This excited embryo population reproduces the high-energy collisions we were looking for given we are optimizing to obtain the most CMF variation. However, the movements of giant planets later on in the formation process can cause more specific dynamical effects. \citet{2020Poon} showed that scattering of outer giant planets can disturb a system of already formed super-Earths and produce this dichotomy. Thus, while our results do not match exactly the eccentricity observations, they are not inconsistent with them as further giant planet motions could be invoked to produce the dichotomy from our current planet systems. 

Our theoretical work calculates the maximum range in CMF from conventional giant impact collisions, and the results point to a maximum of CMF of about 0.6. We also find, in agreement with \cite{2020Scora} and \citet{2021Cambioni}, that this range decreases with increasing planetary mass. Thus, despite optimizing for conditions that produce variation in CMF, conventional giant planet collisions cannot reproduce some of the exoplanet data. For example, Kepler 105c \citep{2016Jontof-Hutter,2017Hadden}, Kepler 406b \citep{2016Morton}, and Kepler 107c \citep{2021Schulze} are all super-Mercuries with high masses and CMF values between 0.6 and 0.8 \citep{Plotnykov2020, 2021Adibekyan}. The composition of these planets is particularly difficult to explain. GJ~367b, on the other hand, is small enough that it may be explained via giant impact collisions. However, it is important to note the large error bars in the CMF inference.

Our work underlines the importance of reducing error bars for mass and radius measurements for these high density super-Earths, so that their CMF can be better known. This will allow for a better estimation of the upper bound of CMF seen in the data.


In addition, it is important to consider that the bias in observations favours large masses for a given radius  \citep{2018Burt,2018Montet}. This translates to favouring high CMF planets. Perhaps their prevalence in the super-Earth data is mostly due to a selection effect, and therefore these planets are rarer than they seem. If this is the case, high-CMF planets as the extremely rare result of collisions is a much better fit to observations than it appears to be currently. 

Future work on both aspects, better mass-radius measurements of super-Mercuries and determining the prevalence of high-CMF planets, is important to make progress. 

As well, future work should consider incorporating the outcomes from more recent collision simulations into the N-body code. New higher-resolution simulations that track vapour production and simulate the collisions of larger bodies have shown that some of the collisional outcomes deviate from those of the \citet{Leinhardt2012} simulations used in this version of \texttt{SyMBA} \citep{2022Reinhardt}. It is important to make use of these newer prescriptions in order to ensure that results are as accurate as possible.

\section{Summary and Conclusions} \label{sec:conc}

We simulated the formation of planets from disks of small, excited embryos with imperfect collisions to form as many high CMF planets as possible. By optimizing conditions for the largest change in CMF during giant impacts, our results correspond to the upper value of CMF in rocky exoplanets. The high excitation of the disks results in elevated impact velocities, and more frequent high-energy collisions. Erosive collisions, one of the most high-energy collisions that occur in our simulations, are 8 times more frequent than in \citet{2020Scora} due to the differences in initial orbital excitations and embryo masses. Despite this, the final distribution of planet CMFs is centered around the initial CMF given to embryos. For planets produced in simulations with no debris mass loss, this distribution is a skewed Gaussian with a small high-CMF tail, with the maximum CMF around 0.6. For planets produced in simulations with $50\%$ debris mass lost due to collisional grinding, this distribution is wider and skewed towards high CMFs, but even so, the maximum CMF is also close to 0.6. 

We find the following conclusions from analyzing the outcomes of these planet formation simulations: 

\begin{itemize}
    \item Highly erosive collisions occur more frequently when embryos have high eccentricities and inclinations, and they can increase a planet's CMF up to 1. However, they tend to occur early on in the planet's formation, when eccentricities are still high. 
    Mass-accreting collisions that tend to have a total averaging effect on planet's CMFs are still the most frequent. Thus, while many planets experience an increase in CMF early on, a combination of debris reaccretion and other collisions will typically increase the planet's mass and reduce the planet's CMF back down to an average that is close to the initial CMF given to the embryos.  
    \item Highly excited systems of embryos produce large quantities of debris, and the dynamical friction from this debris will damp down the eccentricities and inclinations of the remaining embryos. The resulting planet systems have low eccentricities and inclinations similar to planets formed from cold disks. Thus, it is possible to form a cold system from an excited system of embryos. 
    \item Debris mass loss increases the average CMF of planets, but does not strongly increase the maximum CMF planets can achieve. It preferentially removes mantle material from the system, increasing the average CMF of the system. Consequently, there is less mantle debris to be reaccreted by a planet after it suffers a collision, and planet CMFs stay higher. 
    Enough debris remains in the system to keep extremely high CMF planets from forming, however. 
    \item Debris mass loss also produces lower mass planets due to the mass loss from the system. The mass loss affects the average number of collisions each planet undergoes during formation. Planets in simulations with $50\%$ debris loss form after half of the number of collisions as their counterparts in simulations with no debris loss. The differences between systems with debris loss and those without suggests that it may be possible to constrain the fraction of debris loss that occurs during planet formation through planetary system parameters.
    \item Those high CMF planets that do form (between $\sim$ 0.5 - 0.6 CMF) tend to have only a few collisions, most of those either erosive or other high-energy, mantle stripping collisions. This might be a path to explain high CMF planets, but the expectation would be for them to be rare. Given the bias in the data towards observing iron rich planets, this is not an unreasonable prediction. 
    \item High mass planets with high CMF are difficult to form via conventional giant impact collisions. 
    The most massive planet in our simulations with a CMF greater than 0.5 is less than 1 $M_{\oplus}$. Assuming that a rocky planet forms with a CMF close to 0.33, the only way to make it a high CMF planet is to strip much of its mantle off via an erosive collision. In this scenario, if the debris is somehow not reaccreted, much of its mass will be lost in the collision. Thus it will no longer be a high mass planet. Alternatively, it could have numerous collisions throughout its lifetime that continually strip it of mantle material as it accretes mass. However, the results of these simulations show that the process of accreting mass over its lifetime inherently brings its CMF towards the average CMF of the system material.
\end{itemize}

Our results, even when optimized to find the most variation in CMF, fall short when explaining the most iron-enriched planets in the super-Earth sample. Possible explanations are that these planets are exceedingly rare but a product of our bias in detection towards larger masses, and/or that their measured CMF is not as large as it seems but a consequence of large errors in mass and radius. The other possibility is that this work has not completely maximized CMF variation due to collisions, thus, future work will explore the impact of a consistent source of excitation to the embryos, for example, secular perturbation by giant planets.

What we know is that rocky planets, including the ones in our own solar system, form with a diversity of CMFs. Thus, it is an important line of inquiry to investigate how the formation process can create such a range of planetary compositions. This work has demonstrated the importance of debris treatment in the case when embryos have high eccentricities and therefore high energy collisions. It has also highlighted that increasing CMF variation is not as simple as increasing the number or energy of collisions. Instead, it seems that the most favourable way of introducing CMF variation is by a few, energetic collisions near the end of the formation process, which may be made possible by more consistent perturbations throughout formation.

\begin{acknowledgments}
JS and DV are supported by the Natural Sciences and Engineering Research Council of Canada (grant RBPIN-2014-06567).  
This research has made use of the NASA Exoplanet Archive, which is operated by the California Institute of Technology, under contract with the National Aeronautics and Space Administration under the Exoplanet Exploration Program.
We would like to acknowledge that our work was performed on land traditionally inhabited by the Wendat, the Anishnaabeg, Haudenosaunee, Metis and the Mississaugas of the New Credit First Nation.  
\end{acknowledgments}

%

\vspace{5mm}



\facility{Exoplanet Archive}


\software{\texttt{astropy} \citep{astropy:2013,astropy:2018},  
          \texttt{SyMBA} \citep{Duncan1998}}


\appendix
\section{Detailed simulation results}
We display a summary of some of the final parameters of each simulation in Table \ref{tab:results}. The table details the number of planets remaining, the range of CMF and the average CMF of those planets, the angular momentum deficit (AMD) of the planets, and the average number of collisions experienced per planet in that system. The AMD of the planets in each system is calculated using the following equation from \citet{Chambers2001}:
\begin{equation}\label{eq:AMD}
    AMD=\frac{\Sigma_{j} m_{j} \sqrt{ } a_{j}\left[1-\sqrt{\left(1-e_{j}^{2}\right)} \cos i_{j}\right]}{\Sigma_{j} m_{j} \sqrt{ } a_{j}}
\end{equation}
This quantifies the relative eccentricities and inclinations of the remaining bodies. For reference, the AMD of the solar system is 0.0018 when averaged over a million year timespan \citep{Chambers2001}.

\begin{table}
    \caption{Average system parameters resulting from our simulations with and without debris loss. These values are calculated for planets above 0.1 $M_{\oplus}$. CMF$_{avg}$ is the average of the planet CMFs in the system. The AMD is calculated using Equation \ref{eq:AMD}. The average collisions per planet is the average of the number of collisions in each planet's collision history. The first section of the table is runs with no debris loss, followed by runs with $50\%$ debris loss. The name of each simulation contains the information on its initial conditions parameters. For example, 'e01' refers to a simulation with average initial eccentricity of 0.1, 'inc02' refers to a simulation with average initial inclination of 0.2 (rad), and 'alp15' refers to an initial embryo distribution surface density of $\alpha = 1.5$.}
    \centering
    \begin{tabular}{ l | c | c | c | c | c }
         simulation & number of planets & CMF range & CMF$_{avg}$ & AMD & average collisions per planet \\
         \hline
         e01-inc02-alp15-dloss0 & 11 & 0.29 - 0.58 & 0.35 & 0.0041 & 27 \\
         e01-inc02-alp25-dloss0 & 8 & 0.29 - 0.4 & 0.33 & 0.0012 & 38 \\
         e01-inc02-alp25-dloss0\_2 & 8 & 0.33 - 0.36 & 0.33 & 0.00065 & 36 \\
         e01-inc02-alp25-dloss0\_3 & 9 & 0.31 - 0.39 & 0.34 & 0.0023 & 38 \\
         e01-inc05-alp15-dloss0 & 10 & 0.30 - 0.36 & 0.33 & 0.0012 & 17 \\
         e01-inc05-alp25-dloss0 & 10 & 0.3 - 0.41 & 0.34 & 0.0014 & 16 \\
         e03-inc02-alp15-dloss0 & 6 & 0.32 - 0.45 & 0.35 & 0.0012 & 44 \\
         e03-inc02-alp25-dloss0 & 10 & 0.28 - 0.36 & 0.33 & 0.00023 & 18 \\
         e03-inc02-alp25-dloss0\_2 & 9 & 0.30 - 0.41 & 0.34 & 0.00087 & 21 \\
         e03-inc05-alp15-dloss0 & 8 & 0.3 - 0.38 & 0.34 & 0.0035 & 22 \\
         e03-inc05-alp25-dloss0 & 9 & 0.29 - 0.41 & 0.33 & 0.00068 & 16 \\
         e03-inc05-alp25-dloss0\_2 & 10 & 0.31 - 0.37 & 0.34 & 0.0011 & 12 \\
         e03-inc05-alp25-dloss0\_3 & 10 & 0.30 - 0.44 & 0.34 & 0.0018 & 13 \\
         e05-inc02-alp15-dloss0 & 7 & 0.31 - 0. 36 & 0.34 & 0.0011 & 26 \\
         e05-inc02-alp25-dloss0 & 10 & 0.30 - 0.35 & 0.33 & 0.00036 & 15 \\
         e05-inc02-alp25-dloss0\_2 & 9 & 0.30 - 0.42 & 0.34 & 0.00096 & 19 \\
         e05-inc02-alp25-dloss0\_3 & 8 & 0.30 - 0.39 & 0.34 & 0.00048 & 21 \\
         e05-inc05-alp15-dloss0 & 8 & 0.30 - 0.35 & 0.33 & 0.0007 & 19 \\
         e05-inc05-alp25-dloss0 & 8 & 0.30 - 0.37 & 0.33 & 0.001 & 14 \\
         e05-inc05-alp25-dloss0\_2 & 8 & 0.30 - 0.40 & 0.35 & 0.0018 & 14 \\
         \hline
         e01-inc02-alp15-dloss50 & 11 & 0.36 - 0.5 & 0.41 & 0.0024 & 24 \\
         e01-inc02-alp25-dloss50 & 11 & 0.32 - 0.49 & 0.37 & 0.0012 & 17  \\
         e01-inc05-alp15-dloss50 & 12 & 0.32 - 0.6 & 0.41 & 0.0024 & 8 \\
         e01-inc05-alp25-dloss50 & 11 & 0.32 - 0.47 & 0.39 & 0.0028 & 9 \\
         e01-inc05-alp25-dloss50\_2 & 11 & 0.33 - 0.43 & 0.37 & 0.0028 & 5 \\
         e01-inc05-alp25-dloss50\_3 & 12 & 0.31 - 0.51 & 0.38 & 0.0021 & 5 \\
         e03-inc02-alp15-dloss50 & 10 & 0.3 - 0.43 & 0.38 & 0.0027 & 18 \\
         e03-inc02-alp25-dloss50 & 8 & 0.32 - 0.52 & 0.41 & 0.0019 & 16 \\
         e03-inc02-alp25-dloss50\_2 & 11 & 0.35 - 0.44 & 0.39 & 0.0016 & 12 \\
         e03-inc05-alp15-dloss50 & 11 & 0.32 - 0.48 & 0.36 & 0.0083 & 5 \\
         e03-inc05-alp25-dloss50 & 7 & 0.34 - 0.47 & 0.39 & 0.0016 & 8 \\
         e05-inc02-alp15-dloss50 & 8 & 0.33 - 0.43 & 0.38 & 0.00097 & 20 \\
         e05-inc02-alp25-dloss50 & 10 & 0.33 - 0.44 & 0.39 & 0.0016 & 13 \\
         e05-inc02-alp25-dloss50\_2 & 10 & 0.30 - 0.47 & 0.39 & 0.0024 & 9 \\
         e05-inc02-alp25-dloss50\_3 & 8 & 0.32 - 0.53 & 0.38 & 0.0028 & 12 \\
         e05-inc05-alp15-dloss50 & 10 & 0.32 - 0.47 & 0.39 & 0.0018 & 8 \\
         e05-inc05-alp25-dloss50 & 7 & 0.35 - 0.40 & 0.38 & 0.0022 & 6 \\
         e05-inc05-alp25-dloss50\_2 & 8 & 0.34 - 0.46 & 0.39 & 0.0044 & 7 \\
         
    \end{tabular}
    \label{tab:results}
\end{table}

\bibliography{super-paper.bib}

\bibliographystyle{aasjournal}



\end{document}